\documentclass{llncs}
\usepackage{multicol} 
\usepackage{makeidx}  
\usepackage{epsfig,amssymb,array,tabularx,amsmath}
\begin{document}
\def\vm{v_{max}}
\newcommand{\nn}{{\cal N}}
\newcommand{\Pt}{\tilde{\mathbf P}}
\newcommand{\cP}{{\cal P}}
%
\frontmatter          
\pagestyle{headings}  
\addtocmark{} 
\mainmatter              
%
\title{Analytical approaches to cellular automata for traffic flow: 
Approximations and exact solutions}
\titlerunning{Analytical approaches to CA for traffic flow: }  
%
\author{Andreas Schadschneider}
\authorrunning{Andreas Schadschneider}   
%
\tocauthor{Andreas Schadschneider (Universit\"at zu K\"oln)}
\institute{Institut f\"ur Theoretische Physik\\ 
  Universit\"at zu K\"oln\\ 
  D--50937 K\"oln, Germany}

\maketitle              

\begin{abstract}
Cellular automata have turned out to be important tools for the simulation
of traffic flow. They are designed for an efficient impletmentation on
the computer, but hard to treat analytically. Here we discuss several
approaches for an analytical description of the Nagel-Schreckenberg 
(NaSch) model and its variants. These methods yield the exact solution 
for the special case $\vm=1$ of the NaSch model and are good approximations
for higher values of the velocity 
($\vm > 1$). We discuss the validity of these approximations and the
conclusions for the underlying physics that can be drawn from the
success or failure of the approximation.
\end{abstract}
\section{Introduction}

Cellular automata (CA) do not only serve
as simple model systems for the investigation of problems in statistical
mechanics, but they also have numerous applications for 'real' problems
\cite{Wolfram}. Therefore it is not suprising that in recent years CA have 
become quite popular for the simulation of traffic flow (see e.g.\ 
\cite{juelich}).

Here we do not want to discuss the question whether CA are suitable models
for the description of traffic. We try to provide a tool box for the 
determination of the basic properties of the stationary state using 
analytical methods. These methods can also be applied to other CA models.

CA are -- by design -- ideal for large-scale computer simulations.
On the other hand, analytical approaches for the description of
CA are notoriously difficult. This is mainly due to the discreteness
and the use of a parallel updating scheme (which introduces 'non-locality'
into the dynamics). In addition, these models are defined through
dynamical rules (e.g.\ transition probabilities) and usually one does
not have a 'Hamiltonian' description. Therefore standard methods
are not applicable. Furthermore, one has to deal with systems
far from equilibrium which do not satisfy the detailed balance 
condition.

However, there is a need for exact solutions or, at least, for
good approximations. These results as well as other exact statements
may help to greatly reduce the need for computer resources.
The interpretation of simulation data is often difficult because
of the 'numerical noise'. Even in the cases where an exact solution
is not possible, a combination of analytical and numerical methods 
might provide better insights.

Here we will present analytical approaches which allow to solve
the special case $\vm=1$ of the Nagel-Schreckenberg model \cite{NagelS}
exactly. For $\vm > 1$ these methods are only approximations, but they
become exact in certain limits. We will discuss the quality of these
approximations and how relevant information about the underlying physics
can be obtained.


\section{Nagel-Schreckenberg model}

The Nagel-Schreckenberg (NaSch) model \cite{NagelS} is a probabilistic
cellular automaton. Space and time are discrete and hence also
the velocities. The road is divided into cells of length 7.5 m.
Each cell can either be empty or occupied by just one car. The state
of car $j$ ($j=1,\ldots,N$) is characterised by an internal 
parameter $v_j$ ($v_j=0,1,\ldots, \vm$), the momentary velocity of the 
vehicle. In order to obtain the
state of the system at time $t+1$ from the state at time $t$, one has
to apply the following four rules to all cars at the same time
(parallel dynamics):
\begin{description}
\item[R1] Acceleration:\ \ \  $v_j\rightarrow v'_j=v_j+1$\ \ \  
(only if $v_j < \vm$)\\
\item[R2] Braking:\ \ \  $v_j \rightarrow v'_j=d_j$ \ \ \ (for $d_j > v_j$)\\
\item[R3] Randomization:\ \ \  $v'_j\ {\stackrel{p}{\rightarrow}}\
v''_j=v'_j-1$\ \ \ with probability $p$ (if $v'_j>0$)\\
\item[R4] Driving:\ \ \  car $j$ moves $v''_j$ cells.
\end{description}
Here $d_j$ denotes the number of empty cells in front of car $j$, i.e.\ 
the gap or headway. One timestep $t\to t+1$
corresponds to approximately 1 sec in real time \cite{NagelS}. 
\begin{figure}[hb]
\centerline{\psfig{figure=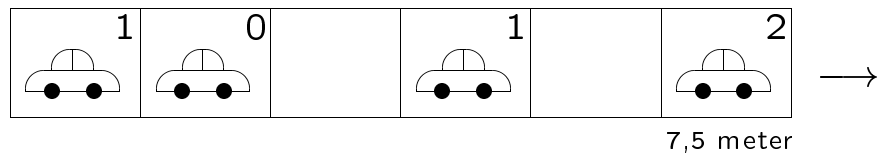,bbllx=60pt,bblly=420pt,bburx=555pt,bbury=535pt,height=2.8cm}}
\caption{Configuration in the Nagel-Schreckenberg model. The number in the 
upper right corner is the velocity of the car.}
\label{fig_conf}
\end{figure}
For simplicity we will only consider periodic boundary conditions so that the 
number of cars is conserved\footnote{A discussion of the effects of open 
boundary conditions can be found in the contribution by
N.\ Rajewsky \cite{niko}. J.\ Krug \cite{joach} and \cite{wolfg}
investigate the effects of disorder.}.
The maximum velocity $\vm$ can be interpreted as a speed limit and is
therefore taken to be identically for all cars.

The four steps have simple interpretations. Step R1 means that every driver
wants to drive as fast as possible or allowed. Step R2 avoids crashes
between the vehicles. The randomization step R3 takes into account several
effects, e.g.\ road conditions (e.g.\ slope, weather) or psychological 
effects (e.g.\ velocity fluctuations in free traffic).
An important consequence of this step is the introduction of an asymmetry 
between acceleration and deceleration, e.g.\ overreactions at braking 
which are important for the occurance of phantom traffic jams. Finally,
step R4 is the actual motion of the vehicles.

The NaSch model is a minimal model in the sense that all four steps R1-R4
are necessary to reproduce the basic properties of real traffic. For more 
complex situations (e.g. 2-lane traffic) additional rules have to be 
formulated. 

In the following sections several methods for an analytical description
of the NaSch model will be presented.


\section{Simple Mean-Field Theory}
\label{sec_MF}

The simplest analytical approach to the NaSch model is a (microscopic)
mean-field (MF) theory. Here one considers the density $c_v(j,t)$ of cars 
with velocity $v$ at site $j$ and time $t$. In the MF approach, correlations 
between sites are completely neglected.

For $\vm=1$ the MF equations for the stationary state ($t\to\infty$)
read \cite{SSNI}:
\begin{eqnarray}
c_0 &=& (c+pd)c,\label{mf1}\\
c_1 &=& qcd \label{mf2}
\end{eqnarray}
with $d=1-c$ and $q=1-p$. The flow is simply given by $f(c)=c_1=qc(1-c)$.

\begin{figure}[t]
\centerline{\psfig{figure=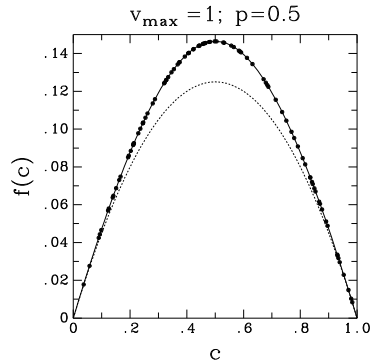,bbllx=40pt,bblly=165pt,bburx=565pt,bbury=680pt,height=5cm}}
\caption{Fundamental diagram for $\vm=1$: Comparison of computer simulations
($\bullet$) with the exact solution (continous line) and the mean-field
result (broken line).}
\label{fig_vmax1}
\end{figure}

For random-sequential dynamics the MF approach is known to be exact 
for $\vm=1$ \cite{NagelS}. For parallel dynamics, however, 
MF theory  underestimates the flow considerably (see Fig.\ \ref{fig_vmax1}).
This shows that correlations are important in this case. These
correlations lead to an increase of the flow. Therefore one expects a 
particle-hole attraction (particle-particle repulsion), i.e.\ the 
probability to find an empty site in front of a car is enhanced 
compared to a completely random configuration. This picture will be 
confirmed by the exact solution.

The MF equations for arbitrary $\vm>1$ are given in \cite{SSNI}.
One finds the same qualitative behaviour, i.e.\ MF theory underestimates
the flow even more than for $\vm=1$.

\section{Garden of Eden States}

\begin{figure}[ht]
\centerline{\psfig{figure=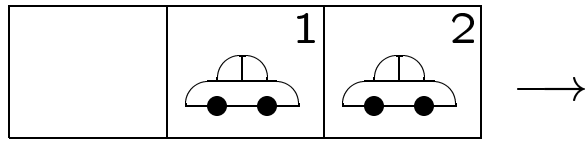,bbllx=150pt,bblly=435pt,bburx=460pt,bbury=520pt,height=2.5cm}}
\caption{A Garden of Eden state for the model with $\vm \geq 2$.}
\label{fig_GoEconf}
\end{figure}

An important effect of the parallel dynamics is the existence of 
configurations which can not be reached dynamically \cite{eden}. These 
states are called Garden of Eden (GoE) states or paradisical states since 
they have no predecessor. An example for a GoE state is given in
Fig.\ \ref{fig_GoEconf}. Note that the velocity is equal to the number 
of cells that the car moved in the previous timestep. For the 
configuration shown in Fig.\ \ref{fig_GoEconf} this implies that the two 
cars must have occupied the same cell before the last 
timestep\footnote{The reader should check 
whether Fig.\ \ref{fig_conf} depicts a GoE state or not !}. 
Since this is forbidden in the NaSch model, the configuration shown can 
never be generated by the dynamics.

The simple mean-field theory presented in the previous section does not
take into account the existence of GoE states. One can therefore hope
that by eliminating all GoE states and applying mean-field theory in
the reduced configuration space (paradisical mean-field, pMF) 
one will find a considerable improvement of the MF results.

For $\vm=1$ all states containing the local configurations
$(0,1)$ or $(1,1)$ are GoE states, i.e.\ the cell behind a car with velocity 1
must be empty. This only affects eq.\ (\ref{mf1}) and 
the equations for pMF theory read:
\begin{eqnarray}
c_0&=&\nn (c_0+pd)c ,\label{pmf10}\\
c_1&=&\nn qcd ,\label{pmf11}
\end{eqnarray}
where the normalization $\nn$ ensures $c_0+c_1=c$ and is given
explicitly by $\nn =\frac{1}{c_0+d}$. 

Solving (\ref{pmf11}) for $c_1$ by using $c_0=c-c_1$, one obtains
$c_1$ as a function of the density $c$. Since the flow is given
by $f(c)=c_1$, 
\begin{equation}
f(c,p) = \frac{1}{2}\left(1-\sqrt{1-4q(1-c)c}\right)
\label{v1flow}
\end{equation}
(with $q=1-p$), we recover the exact solution for the case $\vm=1$
found first in \cite{ss93} using the cluster approximation (see Section
\ref{sec_clust}). Note that the flow depends on the density only via
$c(1-c)$, reflecting the particle-hole symmetry of the NaSch model for 
$\vm=1$.

\begin{figure}[t]
\centerline{\psfig{figure=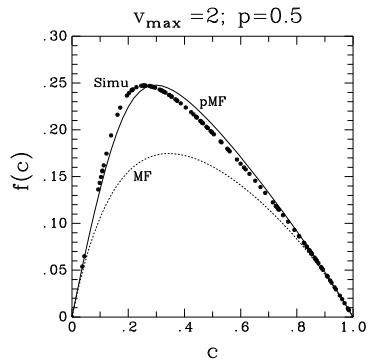,bbllx=55pt,bblly=160pt,bburx=570pt,bbury=680pt,height=5cm}}
\caption{Comparison of the fundamental diagrams obtain from MC simulations
with the results of mean-field theory (MF) and mean-field theory without
GoE states (paradisical mean-field, pMF).}
\label{fig_GoE}
\end{figure}

For $v_{max}=2$ one has to take into account more GoE states \cite{eden}.
pMF is no longer exact, but it still leads to a considerable 
improvement of the MF results (see Fig.\ \ref{fig_GoE}).

It is interesting to note that for random-sequential dynamics MF theory
is exact whereas for parallel dynamics pMF is exact. This suggests that
the important difference between random-sequential and parallel dynamics
is the existence of Garden of Eden states in the latter. 
This is probably not only true for the NaSch model, but rather is a general
property of CA models.


\section{Cluster Approximation}
\label{sec_clust}

The cluster approximation is a systematic improvement of MF theory which
takes into account short-ranged correlations between the cells.
In the $n$-cluster approximation a cluster of $n$ neighbouring cells is
treated exactly. The cluster is then coupled 
to the rest of the system in a self-consistent way. Related approximations
have already been used (under different names) for other models
\cite{kikuchi,guto,benA,cris}.

In order to simplify the description, we choose a slightly different
update-ordering  R2-R3-R4-R1 instead of R1-R2-R3-R4, i.e.\ we look at the 
system after  the acceleration step. Then there are no cars with $v=0$ and
effectively we have to deal with one equation less. In the following
we will use occupation variables $n_j$ where $n_j=0$, if cell $j$ is
empty, and $n_j=v$, if cell $j$ is occupied by a car with velocity $v$.

\begin{figure}[ht]
\centerline{\psfig{figure=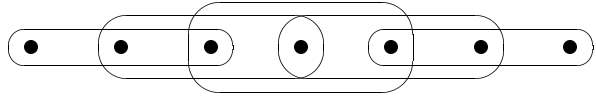,bbllx=100pt,bblly=435pt,bburx=510pt,bbury=530pt,height=2cm}}
\caption{The 3-cluster approximation. Shown are the central 3-cluster and
all neighbouring 3-clusters with a nonvanishing overlap.}
\label{fig_3cluster}
\end{figure}

If we denote the probability to find the system in a configuration
$(n_1,\ldots,n_L)$ by $P(n_1,\ldots,n_L)$ the 1-cluster approximation
means a simple factorization 
\begin{equation}
P(n_1,\ldots,n_L)=\prod_{j=1}^L P(n_j).
\end{equation}
This is nothing but the mean-field theory of section \ref{sec_MF}.
For the 2-cluster approximation one has a factorization of the form
\begin{equation}
P(n_1,\ldots,n_L)\propto P(n_1,n_2)P(n_2,n_3)\cdots 
P(n_{L-1},n_L)P(n_L,n_1).
\label{2clusterapp}
\end{equation}
The 3-cluster approximation is depicted graphically in Fig.\
\ref{fig_3cluster}. 

In general, the master equation leads to $(\vm+1)^n$ nonlinear equations
in $n$-cluster approximation. Although this number can be reduced by using
certain consistency conditions \cite{guto}, a solution is only feasible
for relatively small cluster-sizes \cite{SSNI}. The quality of the
approximation improves with increasing $n$ and for $n\to\infty$ the
$n$-cluster result becomes exact. 

However, for $v_{max}=1$ already the 2-cluster approximation is exact 
\cite{SSNI,ss93}. This is understandable from the results of the
previous section since in this case the 2-cluster approximation is able 
to take into account all GoE states. The 2-cluster probablities for
the stationary state are given explicitly by
\begin{eqnarray}
P(0,0)&=&1-c-P(1,0),\nonumber\\
P(1,1)&=&c-P(1,0),\\
P(1,0)&=&P(0,1)=\frac{1}{2q}\left[1-\sqrt{1-4qc(1-c)}\right],\nonumber
\end{eqnarray}
where again $q=1-p$.
Using $f(c)=qP(1,0)$, one obtains the exact result for the flow (see 
(\ref{v1flow})). Note that the particle-hole attraction is apparent from
these results, since $P(0,1) \ge P(0)P(1)=c(1-c)$.

\begin{figure}[t]
\centerline{\psfig{figure=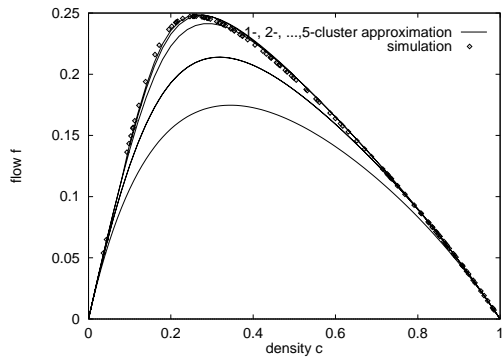,bbllx=40pt,bblly=210pt,bburx=535pt,bbury=560pt,height=5cm}}
\caption{Comparison of simulation results with the $n$-cluster approximation
($n=1,\ldots,5$) for the fundamental diagram for $\vm=2$ and $p=1/2$.}
\label{fig_fund2}
\end{figure}

For $\vm=2$ the fundamental diagrams obtained from the $n$-cluster
approximation ($n=1,\ldots,5$) are compared in Fig.\ \ref{fig_fund2}
with results of Monte Carlo simulations. One can see a rapid convergence,
already for $n=4$ the difference between the simulation and the cluster
result is extremely small.

\subsection{Correlation Length}

As an application of the cluster approximation we will compute 
density-density correlations for $\vm =1$ in the following. Using occupation
numbers $n_j=0,1$ the density-density correlation function is 
defined by
\begin{equation}
\langle n_1 n_r\rangle = {\sum_{\{n_j\}}}'n_1n_r P(n_1,\ldots,n_L)
\label{corrdef}
\end{equation}
where the prime indicates that the sum runs over all states with fixed 
particle number $N=n_1+\cdots +n_L$.

In order to evaluate the sum in (\ref{corrdef}), it is convenient to use
a grand-canonical description. One introduces a fugazity $z$ which controls
the average number of particles $\langle n_j\rangle$ and sums over all
configurations in (\ref{corrdef}). Using the 2-cluster approximation
(\ref{2clusterapp}) -- which is exact for $\vm=1$ -- the correlation 
function is given by
\begin{eqnarray}
\langle n_1 n_r\rangle = \frac{1}{Z_{gc}}{\sum_{\{n_j\}}} n_1n_r 
z^{N} P(n_1,n_2)P(n_2,n_3)\cdots P(n_{L-1},n_L)P(n_L,n_1)
\label{corrgc}
\end{eqnarray}
with $N=\sum_{j=1}^L n_j$ and the normalization
\begin{equation}
Z_{gc}=\sum_{\{n_j\}}z^{N}\prod_{j=1}^L P(n_j,n_{j+1}).
\end{equation}
Introducing the transfer matrix
\begin{equation}
{\Pt}=\begin{pmatrix} 
P(0,0)         & {\sqrt{z}}P(0,1) \\
{\sqrt{z}}P(1,0) & zP(1,1)
\end{pmatrix}
\end{equation}
this can be written succinctly as
\begin{eqnarray}
Z_{gc}&=&{\rm Tr\ } {\Pt}^L\ ,\\
\langle n_1 n_r\rangle &=&\frac{1}{Z_{gc}}\, {\rm Tr}\left({\mathbf Q}
\Pt^{r-1}{\mathbf Q}{\Pt}^{L-r+1}\right).
\label{corrlength}
\end{eqnarray}
with the matrix $Q(n_1,n_2)= n_1\Pt(n_1,n_2)$.

The correlation length $\xi$ can be obtained from the asymptotic 
behaviour ($r\to\infty$) of the correlation function 
\begin{equation}
\langle n_1 n_r\rangle -c^2 \propto e^{-r/\xi}
\end{equation}
where $c=\langle n_j\rangle$ is the (average) density of cars. 
$\xi$ is determined by the ratio 
of the eigenvalues $\lambda_\pm$ of $\Pt$ (with $|\lambda_+| \geq 
|\lambda_-|$):
\begin{equation}
\xi^{-1}=\ln\left|\frac{\lambda_-}{\lambda_+}\right|
\end{equation}
The explicit expression for $\xi$ is rather lenghty. Therefore we just
note that, for fixed $p$, $\xi$ is maximal for $c=1/2$. $\xi(c=1/2)$ diverges 
only for $p\to 0$. In that case one finds that $\xi(c=1/2)\propto p^{-1/2}$.
Monte Carlo simulations show that the same behaviour still occurs (at 
density $c=\frac{1}{\vm+1}$) for $\vm >1$ \cite{eisi,crit}. 
Therefore, the correlation function already gives an indication that 
the system is not critical for $p>0$. The simulations show that there is 
no qualitative difference between the cases $\vm=1$ and $\vm >1$ as far
as the phase transition is concerned \cite{crit}.

The above results demonstrate that, although the 2-cluster approximation
is exact, not all correlation functions are short-ranged or even of
finite range.

\subsection{Further Applications}

We briefly mention other results which have been obtained analytically
using the cluster approximation. In \cite{chowd1} the distributions of
gaps and the distance between jams have been calculated for $\vm=1$ using
the 2-cluster approximation. For the gap distribution one recovers the
(exact) result which will be derived in Sect.\ \ref{seccomf} (see 
eq.\ (\ref{gapdist})). For the calculation of the distribution of gaps between
jams one defines every vehicle with velocity 0 to be jammed. The distance
between two jams is then given by the distance between a car with
velocity 0 and the next car with velocity 0.
In \cite{chowd2} the probability $\cP(t)$ of a time headway $t$ has been
investigated. This quantity is defined in analogy to measurements on
real traffic where a detector registers the time interval between
the passing of consecutive cars.

\section{Car-Oriented Mean-Field Theory}
\label{seccomf}

The car-oriented mean-field (COMF) theory \cite{COMF} is another possibility 
to take into account correlations in an analytical description\footnote{A
similar method is discussed in \cite{interpart}.}.

The central quantitiy in COMF is the probability $P_n(v)$ to find 
exactly $n$ empty cells (i.e.\ a gap of size $n$) in front of a car 
with velocity $v$. In this way certain longer-ranged correlations
are already taken into account. The essence of COMF is now to neglect 
correlations between the gaps.

For $\vm=1$ the system of equations resulting from the master equation
has the following form:
\begin{eqnarray}
P_0(t+1)&=& \bar g(t)\left[P_0(t)+qP_1(t)\right],\nonumber\\
P_1(t+1)&=& g(t)P_0(t) + \left[qg(t)+p\bar g(t)\right]P_1(t) 
          + q\bar g(t)P_2(t),\\
P_n(t+1)&=& pg(t)P_{n-1}(t) + \left[qg(t)+p\bar g(t)\right]P_n(t)
         + q\bar g(t)P_{n+1}(t),\quad (n\geq 2)\nonumber
\label{comfv1}
\end{eqnarray}
where $g(t)=q\sum_{n\geq 1} P_n(t) = q[1-P_0(t)]$ ($\bar g(t)=1-g(t)$)
is the probability that a car moves (does not move) in the next timestep. 
As an example for the derivation of these equations we consider the
equations for $n\geq 2$. Since the velocity difference of two cars is at
most 1, a gap of $n$ cells at time $t+1$ must have evolved form a gap
of length $n-1$, $n$, or $n+1$ in the previous timestep. A headway of $n-1$
cells evolves into a headway of $n$ cells only if the first car moves (with
probability $g(t)$) and the second car brakes in the randomization step
(probability $p$), i.e.\ the total probability for this process is 
$pg(t)P_{n-1}(t)$. Similarly, the headway remains constant only if either
both cars move (probability $qg(t)$) or both cars do not move (probability 
$p{\bar g}(t)$). Finally, the headway is reduced by one, if only the
second car moves (probability $q{\bar g}(t)$).

Although this infinite system of non-linear equations looks more difficult 
than those of the cluster approximation, a solution possible using 
generating functions \cite{COMF}. For $\vm=1$ one finds
\begin{eqnarray}
P_0 &=& \frac{1}{2qc}\left[2qc-1+\sqrt{1-4qc(1-c)}\right],\nonumber\\
P_n &=&\frac{P_0}{p}\left(\frac{p(1-P_0)}{P_0+p(1-P_0)}\right)^n
\quad (n\geq 1),
\label{gapdist}
\end{eqnarray}
and for the flow $f(c,p)=cg$ again the exact solution is reproduced.

For $\vm=2$ one has two coupled systems of the type (\ref{comfv1}), since 
one has to distinguish $P_n(v=1)$ and $P_n(v=2)$ \cite{COMF}. This system 
can also be solved, but as for the other methods, one does not find an 
exact solution.

\begin{figure}[ht]
\centerline{\psfig{figure=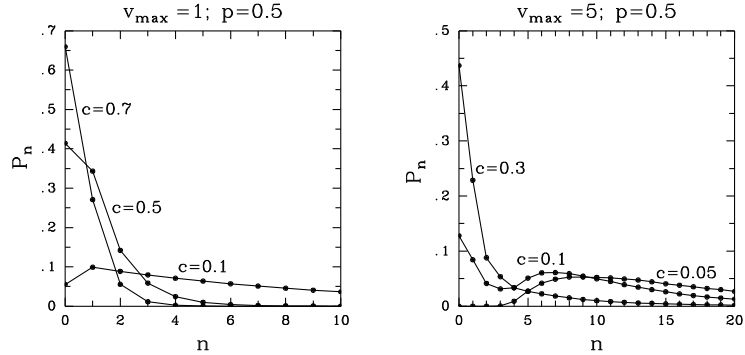,bbllx=40pt,bblly=420pt,bburx=580pt,bbury=685pt,height=5cm}}
\caption{Distribution of headways for different densities for $\vm =1$ 
and $p=0.5$  (left) and $\vm =5$ and $p=0.5$ (right).}
\label{fig_COMF}
\end{figure}

In Fig.\ \ref{fig_COMF} results for the distribution of headways are shown.
For $\vm=1$, the distribution has just one maximum at $n=n_0$, with 
$n_0=1$ for small $c$ and $n=0$ for large $c$. For higher velocities, however,
there is a density regime where the headway distribution exhibits two local
maxima \cite{chowd1}. The maximum at $n=0$ corresponds to jammed cars and
the second maximum corresponds to free flowing cars.


\section{Cluster Approximation vs.\ COMF}

In this section the results of the cluster approximation and COMF in special 
limits are compared. As already mentioned, both methods yield the exact
solution for $\vm=1$. This is related to the fact that paradisical mean-field
is exact in that case and both methods are able to identify all
GoE states. For $\vm=2$ the situation is different. Here all three
methods are only approximative. The quality of COMF is usually between
those of the 2-cluster and 3-cluster approximations.

\begin{figure}[ht]
\centerline{\psfig{figure=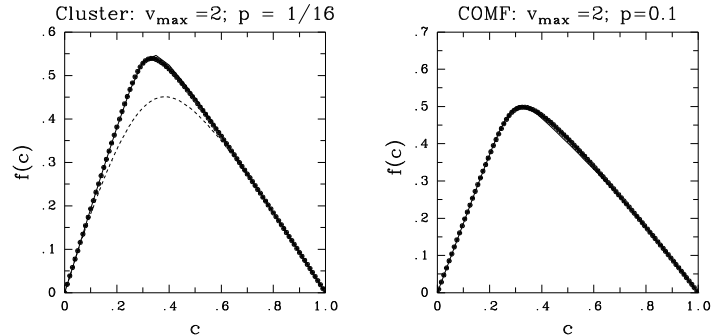,bbllx=45pt,bblly=415pt,bburx=575pt,bbury=695pt,height=5cm}}
\caption{Fundamental diagram for $\vm=2$: Comparison of simulation data 
($\bullet$) with the 2-cluster ($\cdots$) and 3-cluster results ({\bf ---})
for $p=1/16$ (left) and with COMF ({\bf ---}) for $p=0.1$ (right).}
\label{fig_smallp}
\end{figure}

In the limit $p\to 0$, COMF and the 3-cluster approximation become
exact, in contrast to the 2-cluster approximation 
(see Fig.\ \ref{fig_smallp}). Even for values of $p\approx 0.1$
there is an excellent agreement between the fundumental diagrams obtained
analytically and the Monte Carlo simulations. Since 'realistic' values
of $p$ are in the region $p \sim 0.1 - 0.2$ the approximations are
indeed applicable in the relevant parameter regime.

The limit $p\to 1$ is difficult to investigate numerically. Preliminary
results show that COMF not exact in this limit. The cluster
approximations give a much better agreement with Monte Carlo simulations,
but it is not clear yet whether it becomes exact or not.


In order to understand the differences between COMF and the cluster
approximation we calculate the distribution of jam sizes using these
methods. Let $C_n $ be the probability to find a (compact) jam of 
length $n$, i.e.\ $n$ consecutive occupied cells.
$C_n$ is proportional to $P(\underline{0}|1)P(\underline{1}|1)\cdots 
P(\underline{1}|1)P(\underline{1}|1) P(\underline{1}|0)$ in 2-cluster 
approximation. Here we have again used the 
different update-ordering R2-R3-R4-R1. $P(\underline{\sigma}|\tau)$ 
denotes the conditional probability
$P(\sigma,\tau)/(\sum_{\tilde\tau}P(\sigma,\tilde{\tau}))$. Note that
the case $n=1$ has to be treated separately, $C_1 \propto \sum_{v=1}^{\vm}
P(\underline{0}|v) P(\underline{v}|0)$. The number of jams is given
by ${\cal N}_J=\sum_{v=1}^{\vm} P(\underline{v}|0)$ and one obtains
\begin{eqnarray}
C^{(2)}_1 &=& \frac{1}{{\cal N}_J}\sum_{v=1}^{\vm} P(\underline{0}|v) 
    P(\underline{v}|0) \nonumber\\
C^{(2)}_n &=& \frac{1}{{\cal N}_J}\, P(\underline{0}|1) 
    P(\underline{1}|1)^{n-2}P(\underline{1}|1) P(\underline{1}|0)
\qquad (n\geq 2).
\end{eqnarray}
$C^{(2)}_n$ decays exponentially for $n\geq 2$, especially one has
$C_{n+1}\leq C_n$.
For the $m$-cluster approximation one can derive similar expressions. Now
one finds an exponential decay for jam sizes $n\geq m$, i.e.\ 
small clusters may dominate (e.g.\ $C^{(5)}_3 > C^{(5)}_1$ is possible).

In COMF, $C_n$ is proportional to $(1-P_0)P_0\cdots P_0(1-P_0)=(1-P_0)^2
P_0^{n-1}$. The number of jams is propotional to $1-P_0$ so that one finds
\begin{equation}
  C_{n}=(1-P_0)P_0^{n-1}.
\end{equation}
This distribution is purely exponential and $P_n \geq P_{n+1}$ for all $n$.
Therefore COMF is not able to describe clustering or phase separation,
i.e.\ situations where jams with more than one car dominate. Clustering
implies that there are correlations between gaps which are completely
neglected in COMF.

Since for $p\to 0$ COMF is exact, there is no tendency towards
clustering in this limit. On the hand, for $p\to 1$ COMF fails to give
a good description of the phyisics since there is a strong tendency 
towards clustering or phase separation in that limit. Note that for
the (deterministic!) case $p=1$ the difference between $\vm=1$ and 
$\vm>1$ becomes most pronounced, since only for $\vm>1$ metastable states 
exist \cite{crit}. 


\section{Other models}

In this section we discuss briefly two other models which have
been investigated with the methods presented before. Both models are
modified NaSch models. The first one is the $T^2$ model \cite{taka,slow2}. 
It is a NaSch model with a modified acceleration rule. This so-called
'slow-to-start' rule is designed to mimic the delay of a car in restarting,
i.e.\ due to a slow pick-up of the engine or loss of the driver's attention.
We will discuss only the case $\vm=1$. Here the 'slow-to-start' rule
has the important effect of breaking the 'particle-hole' symmetry 
of the NaSch model. 

Explicitly the 'slow-to-start' rule is given by:
\begin{description}
\item[R1'] Acceleration with 'slow-to-start': A standing car will 
always accelerate to $v=1$ if there are at least two empty sites in front;\\
Is there only one empty site, it will accelerate only with probability 
$1-p_t$.
\end{description}

Fig.\ 9 shows the results for the fundamental diagram \cite{slow2}.
Since the 'particle-hole' symmetry is broken, the flow is now maximal
at a density $c_{max} < 1/2$. Another interesting effect can be seen
for larger values of $p_t$: The flow vs.\ density relation has an inflection
point and is no longer convex, i.e.\ the parameter $p_t$ controls 
the curvature of the fundamental diagram (Fig.\ 9). 
This can be seen most clearly for $p_t=1$. Here the flow $f(c)$ 
vanishes for $c \geq 1/2$. This new phase transition to a completely 
jammed state exist only for $p>0$. 

\begin{figure}[hb]
\centerline{\psfig{figure=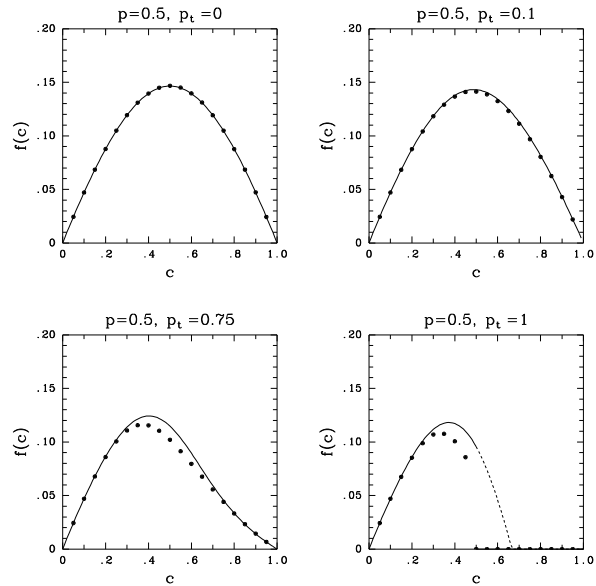,bbllx=70pt,bblly=155pt,bburx=580pt,bbury=700pt,height=8cm}}
\label{fig_t2}
\caption{Fundamental diagram for the T$^2$ model for different values
of $p$ and $p_t$.}
\end{figure}

For $c=1/2$ the stationary state consists
of standing cars with exactly one empty site in between them ($0.0.0.0.
\cdots$ where '$0$' denotes a standing car and '.' an empty site). No car 
can move due to the slow-to-start rule. For larger densities the stationary
state is essentially of the same form, but larger clusters of standing
cars will appear. In Monte Carlo simulations starting from a mega-jam 
the relaxation into this stationary state is extremely slow, even for 
'large' values of $p$, e.g.\ $p=0.5$. 

The COMF equations have two solutions \cite{slow2}. Apart from the solution 
with vanishing flow for $c\geq 1/2$ there exists a second solution with
$f(c)>0$ for $1/2<c<2/3$ (dotted line in Fig.\ 9). 
Therefore one might expect the existence of a
hysteresis effect in this model, which is indeed found in Monte Carlo
simulations\footnote{A more detailed discussion can be found elsewhere 
in these proceedings \cite{robert}.}. This is not surprising since the
slow-to-start rule reduces the outflow from a megajam compared to the
maximum flow of 'free traffic' which seems to be an important factor
for the occurance of hysteresis.

The second model has been introduced by Fukui and Ishibashi \cite{fukui}
and can be considered as a NaSch model for ``aggressive driving''. Again
the rules are identical to the NaSch model, only the acceleration rule
is changed:
\begin{description}
\item[R1''] Acceleration: Every car accelerates to $\vm$. 
\end{description}
For $\vm=1$ this does not change anything, but for higher velocities
it leads to a considerable enhancement of the flow. 

Although the model is less realistic than the NaSch model, it is of 
interest due to its simplicity. An analytical description is much simpler
than that of the NaSch model since after the acceleration step all cars
have the same velocity $\vm$. Therefore the analytical description is
much simpler and might serve as testing ground for new methods \cite{aggr}.

\begin{figure}[hb]
\centerline{\psfig{figure=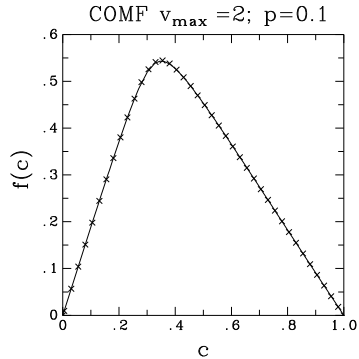,bbllx=80pt,bblly=170pt,bburx=565pt,bbury=690pt,height=5cm}}
\label{fig_raser}
\caption{Fundamental diagram for the Fukui-Ishibashi model: Comparison of
simulation results ($\times$) and COMF ({\bf ---}).}
\end{figure}


\section{Summary}

Although cellular automata are designed for  efficient computer simulation
studies, an analytical description is possible, although difficult.
We have present here four different methods which can be applied
to CA models of traffic flow. The first approach, a simple mean-field
theory for cell occupation numbers, is insufficient since the important 
correlations between neighbouring cells (e.g.\ the particle-hole attraction) 
are neglected. We therefore suggested three different improved mean-field
theories. These approaches take into account certain correlations 
between the cells. The simplest method is the so-called 'paradisical
mean-field' theory which is based on the observation that certain
configurations (Garden of Eden states) can never be generated by the
dynamical rules due to the use of parallel dynamics. The cluster
approximation, on the other hand, treats clusters of a certain size
exactly and couples them in a self-consistent way. Therefore short-ranged
correlations are taken into account properly. In contrast, car-oriented 
mean-field theory is a true mean-field theory, but here one uses a 
different dynamical variable, namely the distance between consecutive
cars. In that way, certain correlations between cells are taken into
account.

All three improved MF theories become exact for $\vm=1$. For larger
values of $\vm$ they are just approximations. In principle, the cluster
approximation and COMF (in combination with a cluster approach) can be
improved systematically. This is, however, very cumbersome.

An interesting observation is that the qualitity of the approximation 
depends strongly on the value of $p$. This indicates that the physics 
changes with $p$, contrary to common believe. Evidence for this
scenario comes from the behaviour of the jam distribution and the
phase transition.

The methods presented here can also be used for other CA models. 
The investigation of the NaSch model has led to a better understanding
of their advantages and limitations so that it is easier to choose
the approach most suitable for a given problem.

The question, how the stationary state is approached, is still
an important open issue. 
Here, generalizations of the methods presented above can also be applied.
This problem is currently under investigation \cite{dynam} and promises
to yield new insights into the physics of the NaSch model.

%
\paragraph{Acknowledgments}

I would like to thank Michael Schreckenberg for our fruitful collaboration
which led to the development of the methods presented here.

\newpage
%
%

\end{document}